# Dynamics and Kinematics of the EUV Wave Event on 6 May 2019

Ramesh Chandra[1,*], P. F. Chen[2], Pooja Devi[1], Reetika Joshi[1] and Y. W. Ni[2]

[1] Department of Physics, DSB Campus, Kumaun University, Nainital 263 001, India;
setiapooja.ps@gmail.com (P.D.); reetikajoshi.ntl@gmail.com (R.J.)

[2] Key Laboratory of Modern Astronomy & Astrophysics, School of Astronomy & Space Science,
Nanjing University, Nanjing 210023, China; chenpf@nju.edu.cn (P.F.C.); y.w.ni@smail.nju.edu.cn (Y.W.N.)

*Correspondence: rchandra.ntl@gmail.com

**Abstract:** We present here the kinematics of the EUV wave associated with a GOES M1.0-class solar flare, which originates in NOAA AR 12740. The event is thoroughly observed with Atmospheric Imaging Assembly (AIA) onboard Solar Dynamics Observatory (SDO) with high spatio-temporal resolutions. This event displays many features of EUV waves, which are very decisive for the understanding of the nature of EUV waves. These features include: a fast-mode wave, a pseudo wave, a slow-mode wave and stationary fronts, probably due to mode conversion. One fast-mode wave also propagates towards the coronal hole situated close to the north pole and the wave speed does not change when it encounters the coronal hole. We intend to provide self-consistent interpretations for all these different features.

1. Introduction

As one of the most energetic phenomena in the solar atmosphere, solar flares are often related to other solar-terrestrial activities, such as coronal mass ejections (CMEs), extreme ultra-violet (EUV) waves, energetic particles, and geomagnetic storms. The source of energy for solar flares is coronal magnetic field, and magnetic reconnection is the widely-accepted mechanism for them [1–4]. According to their association with CMEs, they can be classified into two categories, i.e., eruptive and confined [5]. Eruptive flares are associated with CMEs [6], whereas confined flares are often associated with plasmoid ejections (see the review by [4]). On the other hand, according to the temporal evolution, solar flares are again classified into two types, i.e., long-duration events and impulsive events [2,4]. It was found that long-duration events are normally eruptive flares which are associated with CMEs, whereas impulsive events are often not eruptive in nature (e.g., [2]). However, there are cases where impulsive flares are also associated with CMEs [7,8].

The effects of intense eruptive events are not localized, and might influence the whole solar atmosphere [9]. One of the global effects is EUV waves, which might induce oscillations of remote coronal structures including coronal loops, filaments/prominences and coronal cavities [10–15]. EUV waves are known as several names, for example, EIT waves, coronal waves, coronal propagating fronts (CPFs), and so on [16–18]. For simplicity, here we use EUV waves to describe any wavelike phenomena observed in EUV, bearing in mind that there are probably two types of EUV waves [19]. EUV waves are defined as the arc-shaped bright features (occasionally they can be in an almost complete circular shape) propagating across the majority of the solar disk, which were discovered via the Extreme-Ultraviolet Imaging Telescope (EIT; [20]) about two and half decades ago [16,21]. Since they were discovered via the EIT telescope, they were initially called "EIT waves". EUV waves are associated with solar flares, filament/prominence eruptions and CMEs. Initially, they were attributed to the pressure enhancement by flares, but soon it was realized that they are intimately related to CMEs [22–24].

After the discovery of EUV waves, extensive discussions were ongoing about their kinematics and their nature [13,25–30]. From the propagation characteristics, it seems that a fast-mode wave (or shock wave) is followed by a slower pseudo wave component. The fast-mode wave component is an MHD wave or even shock wave. Its wave nature is manifested as the wave reflection and refraction [31]. Due to their wave nature, people can use them as a tool in coronal seismology. On the other hand, the slower pseudo wave component of EUV waves travels with a speed several times smaller than the fast-mode wave speed in the corona [16,32]. The slower component was observed to stop at the footpoints of the magnetic separatrix [33].

To explain the EUV wave phenomena, several models have been proposed, including the fast-mode wave model, slow-mode soliton model, magnetoacoustic surface gravity waves, current shell, reconnection front model and hybrid model (for details of these models please see the review by Warmuth [9]). The hybrid model proposed by Chen et al. [34,35] is the most suitable model to explain the two components, i.e., a fast-mode wave component and a pseudo wave component. According to this model, the faster outer front is interpreted as a fast-mode MHD wave or shock wave, whereas the inner slower component corresponds to plasma compression due to successive stretching of magnetic field lines which are pushed by an erupting flux rope.

With the SDO observations in unprecedented spatio-temporal resolutions, many new observational features have been discovered. One of the newly-discovered interesting properties of EUV waves is that when the fast-mode component of an EUV wave travels through a separatrix or quasi-separatrix layers (QSLs), a stationary bright front is generated while the fast-mode wave keeps moving with much reduced intensity [36]. A QSL is a generalization of separatrix and is defined as regions where the connectivity of magnetic field lines drastically changes, but remains continuous [37,38]. They often appear in active regions, coronal streamer edges, and the boundary of coronal holes. Such a feature was later confirmed by more observations [29,39]. The generation of the stationary front was explained by Chen et al. [40] to result from mode conversion from the fast-mode wave to the slow-mode wave, which is then trapped in coronal loops, and stops near the footpoints of QSLs.

The objective of this study was to look into the multi-wavelength and multi-viewpoint observations of the EUV wave event on 6 May 2019. The manuscript is organized as follows: In Section 2, the observational datasets are presented. The flare morphology and the EUV wave kinematics are presented in Sections 3 and 4, respectively. The discussion and the summary are given in Section 5.

## 2. Observational Data Sets

The observations of the event are available in different wavelengths from the Atmospheric Imaging Assembly (AIA) [41] on board Solar Dynamics Observatory (SDO) [42] satellite. The cadence and the pixel size of the AIA data are 12 s and 0.6 arcsec, respectively. For our current study, we used AIA 171, 193, and 1600 Å data due to the clarity of the event in these wavelengths. The event was also observed by the STEREO-A spacecraft from a different viewing angle with a cadence of 5 min and a spatial resolution of 1.6 arcsec and the EUV wave event was well visible in 195 Å. Therefore, for the other viewing angle, we used the STEREO-A 195 Å data. In addition, for the magnetic field connectivity, we used the data from the Helioseismic Magnetic Imager (HMI: [43]) on board the SDO satellite. The pixel resolution and cadence of the HMI dataset are 0.5 arcsec and 45 s respectively.

For information on the coronal magnetic configuration, we extrapolate the coronal magnetic field based on the synoptic map of the magnetic field at 06:00 UT on 6 May 2019 using the potential field source surface (PFSS) model. Note that the magnetic field in a synoptic map was usually obtained over a time span of a Carrington period. In this paper we update the radial magnetic field distribution for all the pixels with a heliocentric angle less than $60°$ with the HMI magnetogram at 06:00 UT on 6 May 2019 in order to obtain the instantaneous coronal magnetic configuration.

## 3. Flare Morphology

Here we describe the GOES M1.0-class solar flare on 6 May 2019 from the NOAA active region (AR) 12740. The AR appeared at the east limb on 4 May 2019 as the α-type magnetic configuration and arrived at the west limb on 16 May 2019. The AR produced 15 GOES C-class flares during its disk passage. It was the only AR on the solar surface on 6 May, when it was located at N08E49 with the βγ-type magnetic configuration. The βγ-type magnetic configuration is defined as a sunspot group, which has a general β-type magnetic configuration but contains one (or more) δ sunspots. Out of the fifteen flares during the passage of the AR over the solar disk, seven C-class flares were produced on 6 May 2019. According to GOES satellite observations, the flare under study started at ≈05:05 UT, peaked at ≈05:10 UT and ended at ≈05:12 UT. The GOES X-ray light curves of the flare are shown in Figure 1. As we can see from the light curves, the flare was very impulsive in nature and its total duration was very short (about 8 min). In the decay phase of the flare, GOES observed a small bump at 05:16 UT. The morphology of the flare in AIA 171 and 1600 Å at the three selected times (shown by vertical dashed lines in the GOES light curve) is displayed in the middle and bottom panels of Figure 1. At these wavelengths, the flare started at ≈05:06 UT as a 'C'-shaped structure, as indicated by the arrow in panel (c1). At around 05:10 UT, the 'C'-shaped ribbon changed to two bright kernels marked as '$K_1$' and '$K_2$', respectively. The kernel '$K_1$' was bright until 05:17:26 UT and '$K_2$' remained bright until 05:29:02 UT. At 05:13 UT, the third kernel '$K_3$' appeared and attained its maximum intensity at 05:16 UT, with a life-time of seven minutes. The locations of these three kernels are shown in panels (c2) and (c3) of the figure. The appearance of kernel '$K_3$' was coincident with the small bump observed in the decay phase of GOES X-ray light curve.

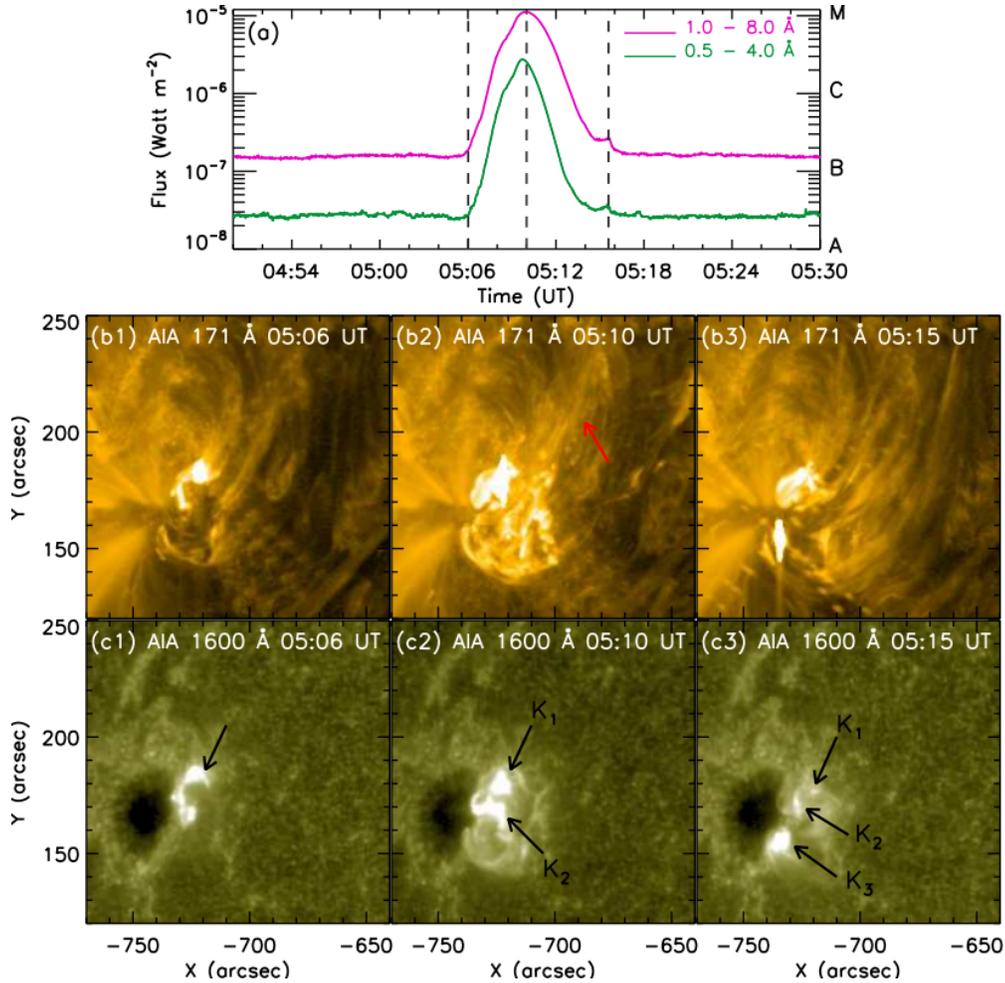

**Figure 1.** (a) Light curves of the flare observed by GOES in two X-ray wavebands. Middle and bottom panels (i.e., panels b1–b3 and c1–c3): Snapshots of the flare evolution at the times indicated by the vertical dashed lines in panel (a). The red arrow in panel (b2) indicates the mass ejected out of the flaring region. $K_1$, $K_2$, and $K_3$ are three flare kernels shown by the black arrows in the bottom panels.

The flare was associated with a very weak CME. According to LASCO CME catalog https://cdaw.gsfc.nasa.gov/CME_list/daily_movies/2019/05/06/, accessed at 6 May 2019 [44]), the speed of the associated CME was ≈240 km s$^{-1}$ and its angular width was 53°.

## 4. EUV Wave Kinematics

An interesting aspect of this CME/flare event is the associated EUV waves. Here we present the EUV wave kinematics. After the flare onset at ≈05:04 UT, EUV waves were visible travelling away from the eruption source region, which was in the AR 12740. The evolution of the wave in the AIA 193 Å base-difference images are depicted in Figure 2. From the figure, it is inferred that initially the EUV wave phenomenon was manifested as a bright arc-shaped front and it was bright until 05:20 UT. As time progressed, the wave became fainter and fainter, and the major part of the wave moved towards the west direction. The wave was followed by dimmings, clearly visible in the AIA 193 Å basedifference images. We also inspected the evolution of the wave in the STEREO-A 195 Å images. Figure 3 presents the wave evolution in the STEREO-A 195 Å base-difference images. The wave first appeared at 05:10:30 UT when it was 20000 away from the origin site. It propagated mainly along the northwest direction, which is different from the AIA observations. The EUV wave did not appear earlier because of the low cadence of the STEREO observations, i.e., 5 min. For a clearer view of this EUV wave event, we refer the readers to the supplemented AIA and STEREO-A animations.

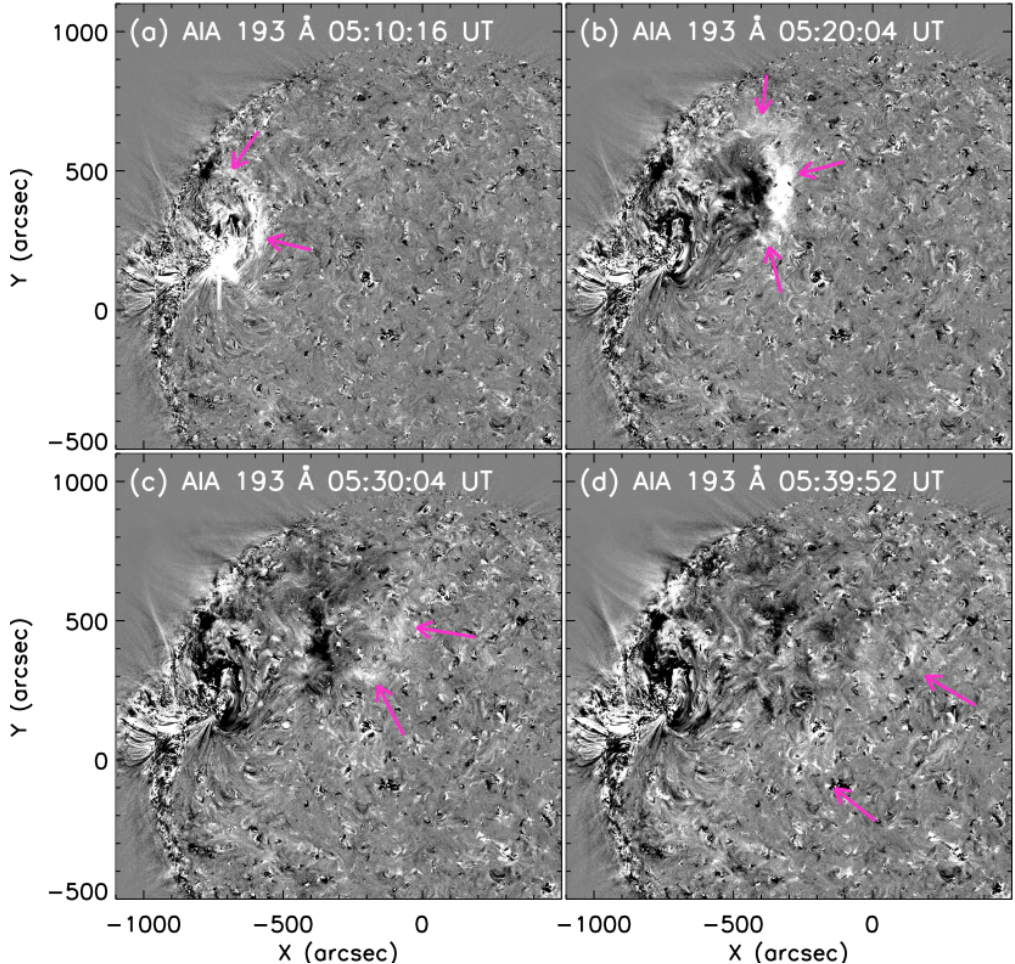

**Figure 2**. Evolution of the base difference images in AIA 193 Å revealing the wave propagation. The base image is selected at 04:50:04 UT. The pink arrows indicate the propagation of the wave.

In order to understand the kinematics of the EUV wave event, we present here the kinematic properties of the EUV wave event in different directions. For this purpose, we first selected a slice starting from the source region, i.e., $S_1$ as indicated in panel (a) of Figure 4. The slice was towards the northwest direction, ending inside the coronal hole region with a distance of ≈700". The boundary of the coronal hole is marked by the blue line in panel (a) of the figure. The wave propagation along this selected direction is plotted in panel (b) of Figure 4. According to this plot, the onset of the EUV wave was at ≈05:09 UT, which is about 100" from the flare source location. Since the solar flare peaked at 05:10 UT when the EUV wave had already propagated 150" away from the flaring loops, our results once again imply that the EUV wave is not driven by the pressure pulse in the solar flare, and it should be driven by the CME, as claimed by Chen [23]. The bright ridge in Figure 4b means an EUV wave was propagating outward along the slice. In order to calculate the front speed, we fit the bright ridge with a straight line. The computed speed was ≈640 ± 8 km s$^{-1}$. One interesting feature in Figure 4b is that the front speed changes little across the boundary of the coronal hole.

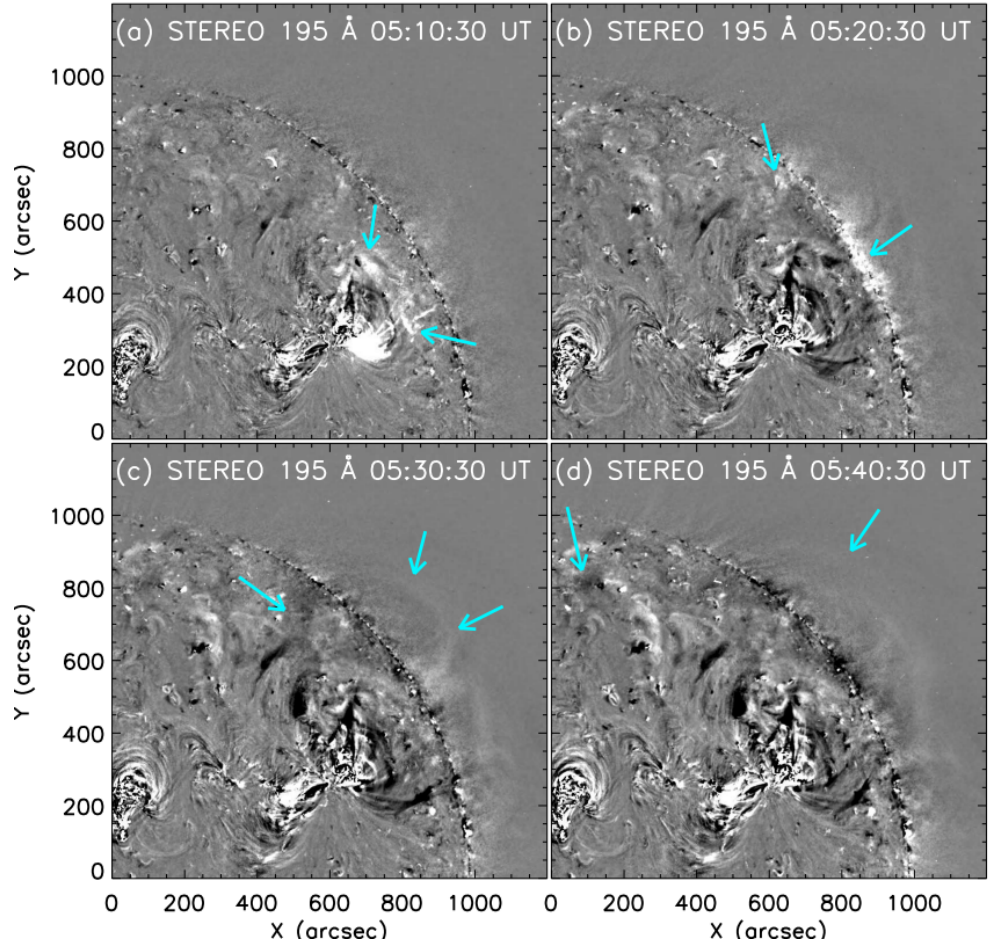

**Figure 3**. Evolution of the base difference images in STEREO-A 195 Å showing the wave propagation. The base image is selected at 04:50:30 UT. The cyan arrows indicate the propagation of the wave.

Since EUV waves are strongly anisotropic, we selected another two slices to see the wave kinematics along the corresponding directions, which are marked as 'S$_2$' and 'S$_3$' in panel (a) of Figure 4 as indicated by the cyan and green arrows. The corresponding time–distance diagrams are plotted in panels (c) and (d) of the same figure. Along the slice S$_2$, it is seen that a bright EUV wave front propagated out with a speed of 802 ± 15 km s$^{-1}$, which then decelerated to 350 ± 3 km s$^{-1}$. The propagation speed is typical for fast-mode EUV waves. On the other hand, we can see the propagation of patchy EUV fronts. The apparent velocity of the patchy fronts is $208 \pm 10$ km s$^{-1}$, about four times slower than the fast-mode EUV wave. Along the slice S$_3$, the onset of the EUV wave in this direction was at the same time as in the S1 direction, i.e., 05:09 UT. However, the evolution of its wave fronts is much different from slice 'S$_1$'. It is seen that from 05:09 to 05:13 UT, only one wave was observed to propagate to a distance of ~400″, which had a speed of $800 \pm 12$ km s$^{-1}$. Interestingly, after 05:13 UT, the EUV wave bifurcated into two fronts. The speeds of these two fronts were markedly smaller than those in the initial phase. The outer front travelled with a speed of about $250 \pm 3$ km s$^{-1}$. It kept almost a constant speed until 05:27 UT, when another stationary front was generated at a distance of 700″. Since such a place corresponds to another QSL, it is expected that wave mode conversion happened here as well. The inner front travelled with a speed of $160 \pm 2$ km s$^{-1}$ until 05:23 UT. After that, this inner wave front nearly stopped at the footpoints of the magnetic field lines.

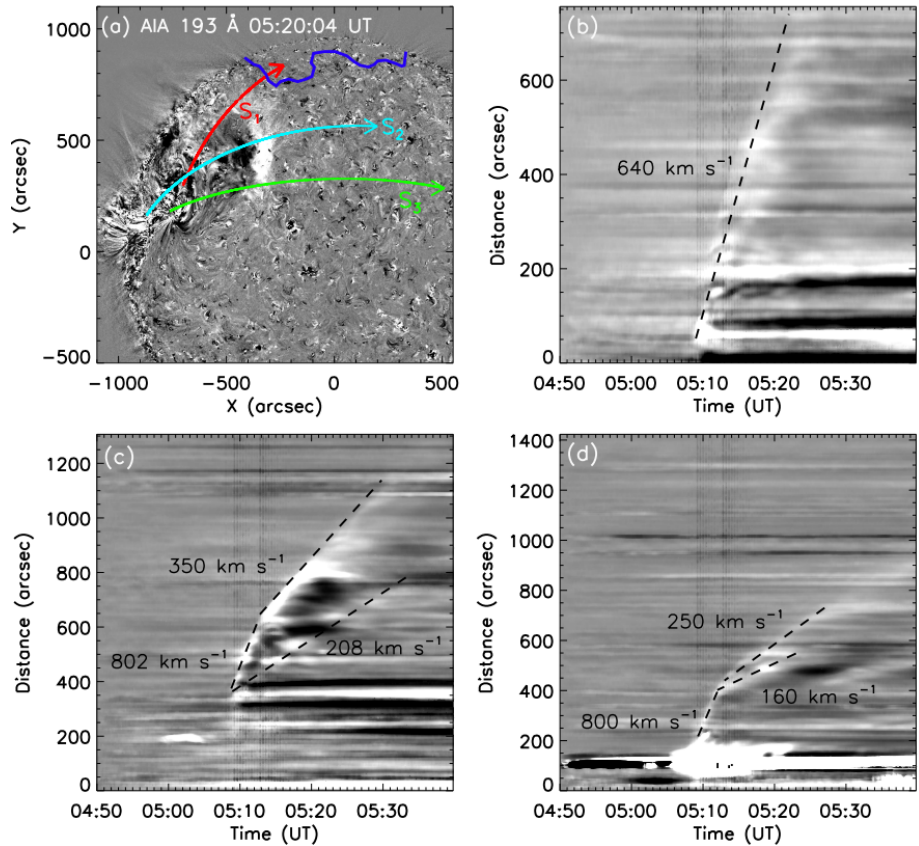

**Figure 4.** Panel (a): AIA base-difference image at 05:20:04 UT, where the blue line marks the boundary of the polar coronal hole. Three slices (S$_1$, S$_2$ and S$_3$) are selected in order to plot the time-distance diagrams of the AIA 193 Å intensity, which are displayed in panels (b–d), respectively.

## 5. Summary and Discussions

In this paper, we presented an EUV wave event accompanied by a GOES M1.0-class solar flare. The main points of this study are as follows:

- Despite the flare being impulsive and compact in nature, it was associated with a weak CME. Similar events were studied before [7,45–47].

- The EUV wave event is strongly anisotropic. We selected three slices along different directions. It is seen that along the northwest direction $S_1$, only one wave was detected, and the propagation speed was $640 \pm 8$ km s$^{-1}$. Along the west direction $S_3$, only one wave was detected initially, with a speed of $800 \pm 12$ km s$^{-1}$. However, it split into two fronts with much smaller speeds, i.e., $250 \pm 3$ km s$^{-1}$ and $160 \pm 2$ km s$^{-1}$, respectively. Along the direction $S_2$, which is between $S_1$ and $S_3$, we detected two EUV waves, one is the fast-mode wave, and the other one is a pseudo-wave with patchy fronts.

In our case, we observed that the interaction between the EUV wave and the coronal hole did not change the front speed, and the EUV wave propagated easily through the boundary of the coronal hole. In some events, it was found that when an EUV wave interacts with a coronal hole, the wave experiences reflection [27,31,48,49] in addition to transmission Olmedo et al. [50]. The reflection and transmission of the wave at the coronal hole boundary was represented in the simulations of Schmidt and Ofman [51]. Their simulations presented a 3-dimensional MHD model to explain the EUV wave interaction with coronal holes, which found that the impact of the EUV wave generates the resonant oscillation of the coronal hole, and the resonant oscillation would generate secondary EUV waves. Recently, Piantschitsch et al. [52] presented a 2.5-dimensional simulation of fast-mode EUV wave interaction with a coronal hole and found the EUV wave can transmit through the coronal hole. It is noted that if the fast-mode wave speed changes rapidly across the boundary of a coronal hole, it is expected to see both the reflection and transmission of an impinging EUV wave. In our case, however, no reflection was detected, and only transmission was seen. Two possible reasons may account for such a feature: One is that the impinging fast-mode wave is already too weak, and any reflected wave is too faint to be detectable; The other is that the fast-mode wave speed across the boundary of this coronal hole changes little. In order to find which reason is more plausible, we plot the distribution of the magnetic field strength at the altitude of $0.1 R_\odot$ along the slice $S_1$ in Figure 5, where the magnetic field strength is taken from the PFSS extrapolation. It is seen that the magnetic field decreases from the source active region to the quiet region, but then gradually increases toward the coronal hole near the far end of the slice. Considering that the plasma density in the coronal hole is significantly reduced compared to that in quiet regions, the fast-mode wave speed across the boundary of the coronal hole should increase even more drastically than in Figure 5. Therefore, it seems that the first reason is more plausible, i.e., the reflected wave is too weak to be detected.

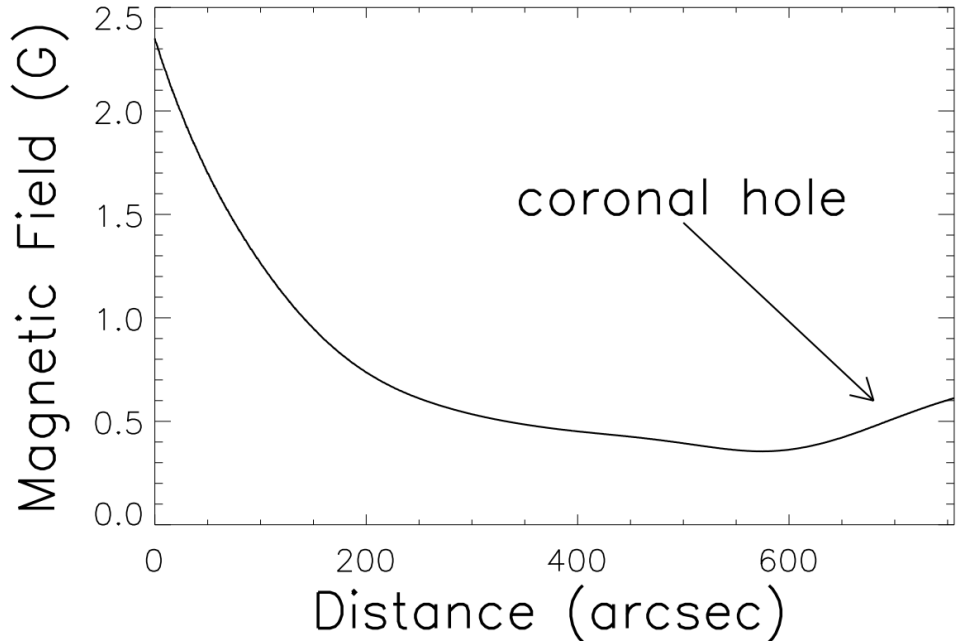

**Figure 5.** Distribution of the magnetic field strength at the altitude of $0.1R_\odot$ along the slice $S_1$.

Another interesting feature of this event is that the initial EUV wave was split into two fronts in the west direction at a distance of 400″ from the source region. The initial EUV wave had a speed of $800 \pm 12$ km s$^{-1}$, a typical value of the coronal fast-mode wave speed. Therefore, it should be a fast-mode wave. Regarding the two fronts, the faster one had a speed of only $250 \pm 3$ km s$^{-1}$, and the slower one had a speed of $160 \pm 2$ km s$^{-1}$. There is no reason for a fast-mode MHD wave to disappear suddenly. Therefore, it is speculated that the faster wave with a speed of $250 \pm 3$ km s$^{-1}$ should be a fast-mode MHD wave as well. However, the speed was decreased suddenly from $800 \pm 12$ km s$^{-1}$ to $250 \pm 3$ km s$^{-1}$, which means that the magnetic field dropped suddenly at the distance of 400″ from the source active region. Figure 4 also indicates that the fast-mode EUV wave generated another stationary wave front at a distance of 700″ from the source active region at 05:27 UT, whereas the slower EUV wave propagated out with a speed of $160 \pm 2$ km s$^{-1}$, and then stopped at a distance of 500″ from the source active region since 05:23 UT.

In order to understand why the fast-mode speed dropped suddenly and what the nature of the slower EUV wave after wave splitting is, we display the extrapolated coronal magnetic configuration in Figure 6, where the blue lines correspond to the magnetic field lines, and the red star marks the location of the source active region. The short green bar indicates the location of EUV wave splitting, and the short yellow bar indicates the location where the fast-mode EUV wave generated a stationary wave front. It is seen that the wave splitting happened in front of a magnetic separatrix. According to Chen et al. [40], such a place is a favorite site for mode conversion since it is above the neutral line between opposite magnetic polarities. Therefore, we can speculate that the slower EUV wave after 05:13 UT was a slow-mode MHD wave.

Such a speculation is further supported by two facts, i.e., this slower EUV wave had a speed of $\sim 160 \pm 2$ km s$^{-1}$, which is the typical sound speed in the corona, and this slower EUV wave stopped since 05:23 UT when it reached the location of a magnetic separatrix. These observational features are consistent with the picture of a slow-mode MHD wave propagating along the magnetic field lines at the edge of a magnetic separatrix or a QSL.

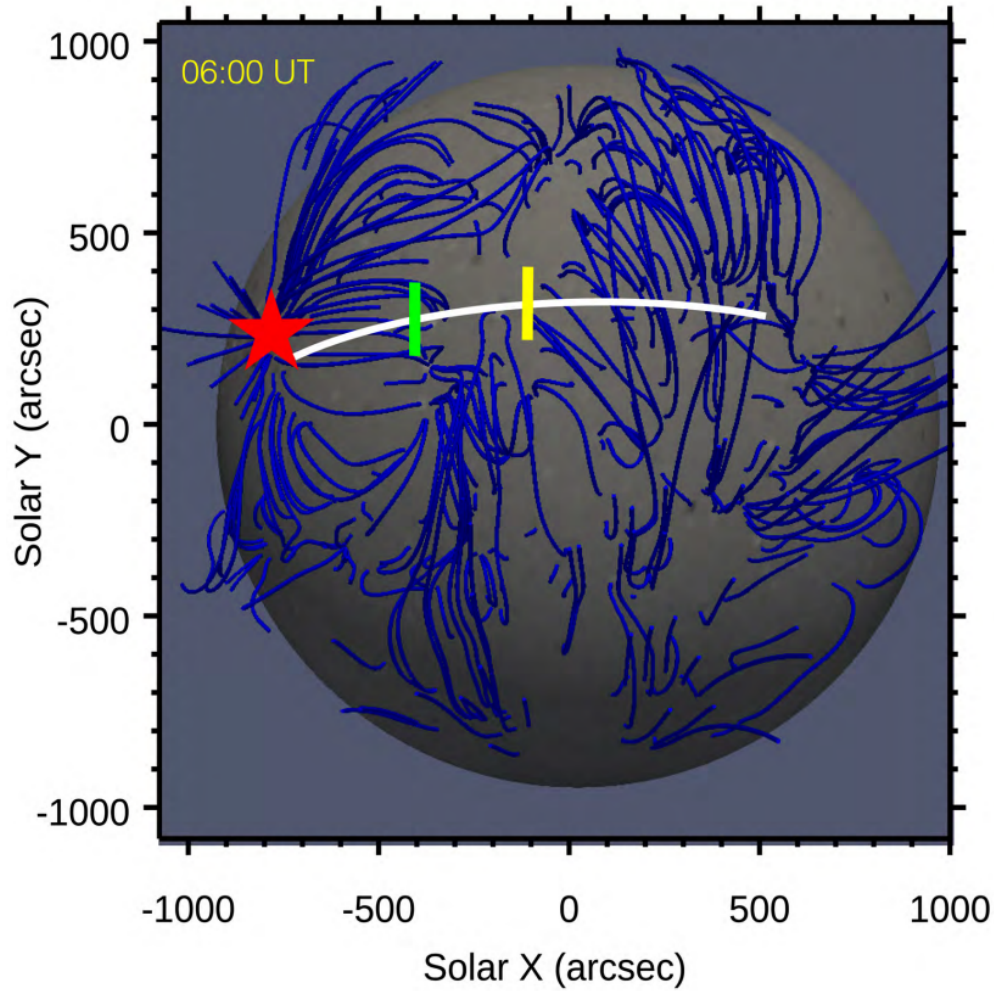

**Figure 6.** Extrapolated coronal magnetic field distribution on 6 May 2019 at 06:00 UT derived via the potential field source surface (PFSS) model. The green bar indicates the location where a single wave bifurcated into two waves at 05:12 UT as indicated by Figure 4d, and the yellow bar marks the location where the fast-mode outer EUV wave changes to a stationary front at 05:27 UT in Figure 4d. The active region is represented by the red star. The white line is the slice $S_3$ drawn in Figure 4a for the time-distance analysis of the wave propagation.

Figure 6 also indicates that the stationary front generated by the faster EUV wave corresponds to the location of another magnetic separatrix or QSL, which once again is consistent with the theoretical model of Chen et al. [40], i.e., a fast-mode MHD wave might be converted to a slow-mode MHD wave which then stops at the footpoint of a magnetic separatrix or QSL.

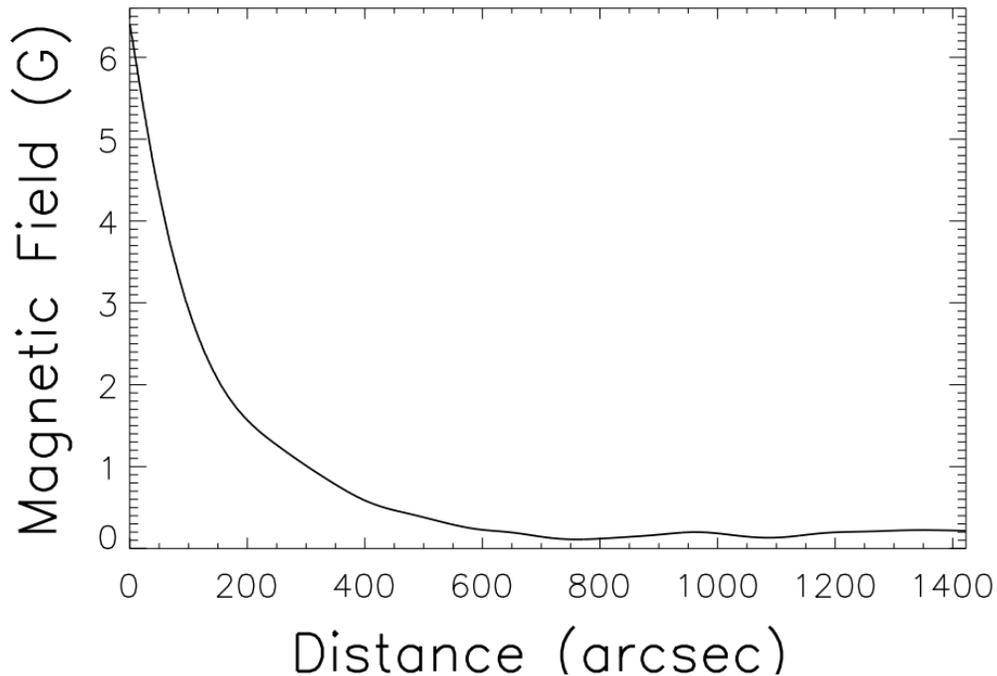

Figure 7. Distribution of the magnetic field strength at the altitude of $0.1R_\odot$ along the slice $S_3$.

One puzzling feature of this EUV wave event is that the fast-mode wave speed dropped drastically at a distance of 400″ from the source active region. From the extrapolated coronal magnetic configuration, we can see in Figure 6 that from the green bar to the yellow bar and beyond, the area was covered by several magnetic systems with alternating positive and negative magnetic polarities. In such a multi-polarity region, the coronal magnetic field generally decays rapidly with height. This is probably why the fast-mode EUV wave was decelerating from $800 \pm 12$ km s$^{-1}$ to $250 \pm 3$ km s$^{-1}$ after the distance of 400″ from the source active region. We tentatively propose a concept of "coronal magnetic holes" to describe this type of area where the coronal magnetic field is significantly lower than surroundings due to its multi-polarity nature. To confirm it, we plot the distribution of the coronal magnetic field strength at the altitude of $0.1R_\odot$ along the slice $S_3$. Figure 7 indeed shows that the magnetic field decreases drastically from more than 6 G near the proximity of the source active region to 0.2 G in the quiet region to the west of the source active region, which is much weaker than the quiet region to the north. Such a drastic deceleration of the fast-mode EUV wave has not yet been shown by any numerical simulations before, so it is worth investigating further via MHD numerical simulations.

It is noticed that along both slices $S_1$ and $S_3$ we did not detect pseudo waves, and only along the slice $S_2$ we detected both the fast-mode EUV wave and the pseudo wave, and the pseudo wave is due to successive appearance of patchy EUV fronts, as discovered by Guo et al. [13]. However, different from their explanation, we argue that the pseudo wave with patchy fronts can still be accommodated in the magnetic field line stretching model proposed by Chen et al. [34,35].

While the fast-mode wave and pseudo-wave have completely different formation mechanisms, both of them present anisotropy on the solar disk. The anisotropy of the fast-mode EUV wave is due to the inhomogeneity of the fast-mode wave speed in the corona. The slower pseudo-wave is even more anisotropic, which is evident only along the slice $S_2$. According to the magnetic field line stretching model proposed by Chen et al. [34,35], pseudo waves propagate only in the regions where the magnetic field lines straddle over the eruption source region.

**Author Contributions**: R.C. contributed to the data analysis and the writing of the draft. P.F.C. contributed to the interpretation and the writing of the manuscript. P.D. contributed to the discussions, and R.J. contributed to the data analysis and Y.W.N. made the coronal magnetic field extrapolation. All authors have read and agreed to the published version of the manuscript.

**Funding:** This work was financially supported by the National Key Research and Development Program of China (2020YFC2201200) and NSFC (11961131002).

**Data Availability Statement:** The datasets analyzed during the current study are available at http://jsoc.stanford.edu/ and https://cdaw.gsfc.nasa.gov/stereo/.

**Acknowledgments:** We acknowledge NASA's open data policy in using the SDO and STEREO data. We would like to thank the developers of Solar Software IDL (SSWIDL) to make it freely available. We thank the reviewers for their comments and suggestions. P.F.C. is supported by the National Key Research and Development Program of China (2020YFC2201200) and NSFC (11961131002 and 11533005). P.D. thank the CSIR, New Delhi for providing research fellowship.

**Conflicts of Interest:** The authors declare no conflict of interest.

**References:**

1. Parker, E.N. The Solar-Flare Phenomenon and the Theory of Reconnection and Annihiliation of Magnetic Fields. Astrophys. J. Suppl. 1963, 8, 177. doi:10.1086/190087.
2. Benz, A.O. Flare Observations. Living Rev. Sol. Phys. 2008, 5, 1. doi:10.12942/lrsp-2008-1.
3. Fletcher, L.; Dennis, B.R.; Hudson, H.S.; Krucker, S.; Phillips, K.; Veronig, A.; Battaglia, M.; Bone, L.; Caspi, A.; Chen, Q.; et al. An Observational Overview of Solar Flares. Space Sci. Rev. 2011, 159, 19–106. doi:10.1007/s11214-010-9701-8.
4. Shibata, K.; Magara, T. Solar Flares: Magnetohydrodynamic Processes. Living Rev. Sol. Phys. 2011, 8, 6. doi:10.12942/lrsp-2011-6.
5. Svestka, Z. On the varieties of solar flares. In The Lower Atmosphere of Solar Flares; Neidig, D.F., Machado, Proceedings of the Solar Maximum Mission Symposium, Sunspot, NM, National Solar Observatory; M.E., Eds.; 1986, pp. 332–355. https://ui.adsabs.harvard.edu/abs/1986lasf.conf..332S.
6. Chen, P.F. Coronal Mass Ejections: Models and Their Observational Basis. Living Rev. Sol. Phys. 2011, 8, 1. doi:10.12942/lrsp2011-1.
7. Uddin, W.; Jain, R.; Yoshimura, K.; Chandra, R.; Sakao, T.; Kosugi, T.; Joshi, A.; Despande, M.R. Multi-Wavelength Observations of an Unusual Impulsive Flare Associated with Cme. Sol. Phys. 2004, 225, 325–336, doi:10.1007/s11207-004-5002-2.

8. Chandra, R.; Jain, R.; Uddin, W.; Yoshimura, K.; Kosugi, T.; Sakao, T.; Joshi, A.; Deshpande, M.R. Energetics and Dynamics of an Impulsive Flare on 10 March 2001. Sol. Phys. 2006, 239, 239–256, doi:10.1007/s11207-006-0098-1.
9. Warmuth, A. Large-scale Globally Propagating Coronal Waves. Living Rev. Sol. Phys. 2015, 12, 3. doi:10.1007/lrsp-2015-3.
10. Asai, A.; Ishii, T.T.; Isobe, H.; Kitai, R.; Ichimoto, K.; UeNo, S.; Nagata, S.; Morita, S.; Nishida, K.; Shiota, D.; et al. First Simultaneous Observation of an Hα Moreton Wave, EUV Wave, and Filament/Prominence Oscillations. Astrophys. J. Lett. 2012, 745, L18, doi:10.1088/2041-8205/745/2/L18.
11. Kumar, P.; Cho, K.S. Simultaneous EUV and radio observations of bidirectional plasmoids ejection during magnetic reconnection. Astron. Astrophys. 2013, 557, A115, doi:10.1051/0004-6361/201220999.
12. Shen, Y.; Liu, Y.D.; Chen, P.F.; Ichimoto, K. Simultaneous Transverse Oscillations of a Prominence and a Filament and Longitudinal Oscillation of Another Filament Induced by a Single Shock Wave. Astrophys. J. 2014, 795, 130, doi:10.1088/0004-637X/795/2/130.
13. Guo, Y.; Ding, M.D.; Chen, P.F. Slow Patchy Extreme-ultraviolet Propagating Fronts Associated with Fast Coronal Magnetoacoustic Waves in Solar Eruptions. Astrophys. J. Suppl. 2015, 219, 36. doi:10.1088/0067-0049/219/2/36.
14. Srivastava, A.K.; Singh, T.; Ofman, L.; Dwivedi, B.N. Inference of magnetic field in the coronal streamer invoking kink wave motions generated by multiple EUV waves. Mon. Not. R. Astron. Soc. 2016, 463, 1409–1415, doi:10.1093/mnras/stw2017.
15. Zhang, Q.M.; Ji, H.S. Vertical Oscillation of a Coronal Cavity Triggered by an EUV Wave. Astrophys. J. 2018, 860, 113, doi:10.3847/1538-4357/aac37e.
16. Thompson, B.J.; Plunkett, S.P.; Gurman, J.B.; Newmark, J.S.; St. Cyr, O.C.; Michels, D.J. SOHO/EIT observations of an Earthdirected coronal mass ejection on 12 May 1997. Geophys. Res. Lett. 1998, 25, 2465–2468. doi:10.1029/98GL50429.
17. Hudson, H.S.; Karlický, M. Global Coronal Waves: Implications for HESSI. High Energy Solar Physics Workshop—Anticipating HESSI; Ramaty, R., Mandzhavidze, N., Eds.; Astronomical Society of the Pacific Conference Series; 2000; Volume 206, p. 268. https://ui.adsabs.harvard.edu/abs/2000ASPC..206..268H.
18. Nitta, N.V.; Schrijver, C.J.; Title, A.M.; Liu, W. Large-scale Coronal Propagating Fronts in Solar Eruptions as Observed by the Atmospheric Imaging Assembly on Board the Solar Dynamics Observatory—An Ensemble Study. Astrophys. J. 2013, 776, 58, doi:10.1088/0004-637X/776/1/58.
19. Chen, P.F. Global Coronal Waves. In Washington DC American Geophysical Union Geophysical Monograph Series; 2016; Volume 216, pp. 381–394. doi:10.1002/9781119055006.ch22.
20. Delaboudinière, J.P.; Artzner, G.E.; Brunaud, J.; Gabriel, A.H.; Hochedez, J.F.; Millier, F.; Song, X.Y.; Au, B.; Dere, K.P.; Howard, R.A.; et al. EIT: Extreme-Ultraviolet Imaging Telescope for the SOHO Mission. Sol. Phys. 1995, 162, 291–312. doi:10.1007/BF00733432.
21. Moses, D.; Clette, F.; Delaboudinière, J.P.; Artzner, G.E.; Bougnet, M.; Brunaud, J.; Carabetian, C.; Gabriel, A.H.; Hochedez, J.F.; Millier, F.; et al. EIT Observations of the Extreme Ultraviolet Sun. Sol. Phys. 1997, 175, 571–599. doi:10.1023/A:1004902913117.
22. Cliver, E.W.; Laurenza, M.; Storini, M.; Thompson, B.J. On the Origins of Solar EIT Waves. Astrophys. J. 2005, 631, 604–611. doi:10.1086/432250.
23. Chen, P.F. The Relation between EIT Waves and Solar Flares. Astrophys. J. Lett. 2006, 641, L153–L156. doi:10.1086/503868.
24. Patsourakos, S.; Vourlidas, A. On the Nature and Genesis of EUV Waves: A Synthesis of Observations from SOHO, STEREO, SDO, and Hinode (Invited Review). Sol. Phys. 2012, 281, 187–222, doi:10.1007/s11207-012-9988-6.

25. White, S.M.; Thompson, B.J. High-Cadence Radio Observations of an EIT Wave. Astrophys. J. Lett. 2005, 620, L63–L66. doi:10.1086/428428.
26. Veronig, A.M.; Temmer, M.; Vršnak, B.; Thalmann, J.K. Interaction of a Moreton/EIT Wave and a Coronal Hole. Astrophys. J. 2006, 647, 1466–1471, doi:10.1086/505456.
27. Long, D.M.; Gallagher, P.T.; McAteer, R.T.J.; Bloomfield, D.S. The Kinematics of a Globally Propagating Disturbance in the Solar Corona. Astrophys. J. Lett. 2008, 680, L81, doi:10.1086/589742.
28. Warmuth, A.; Mann, G. Kinematical evidence for physically different classes of large-scale coronal EUV waves. Astron. Astrophys. 2011, 532, A151. doi:10.1051/0004-6361/201116685.
29. Chandra, R.; Chen, P.F.; Joshi, R.; Joshi, B.; Schmieder, B. Observations of Two Successive EUV Waves and Their Mode Conversion. Astrophys. J. 2018, 863, 101. doi:10.3847/1538-4357/aad097.
30. Chandra, R.; Chen, P.F.; Devi, P.; Joshi, R.; Schmieder, B.; Moon, Y.J.; Uddin, W. Fine Structures of an EUV Wave Event from Multi-viewpoint Observations. Astrophys. J. 2021, 919, 9, doi:10.3847/1538-4357/ac1077.
31. Gopalswamy, N.; Yashiro, S.; Temmer, M.; Davila, J.; Thompson, W.T.; Jones, S.; McAteer, R.T.J.; Wuelser, J.P.; Freeland, S.; Howard, R.A. EUV Wave Reflection from a Coronal Hole. Astrophys. J. Lett. 2009, 691, L123–L127. doi:10.1088/0004-637X/691/2/L123.
32. Zhukov, A.N.; Rodriguez, L.; de Patoul, J. STEREO/SECCHI Observations on 8 December 2007: Evidence Against the Wave Hypothesis of the EIT Wave Origin. Sol. Phys. 2009, 259, 73–85. doi:10.1007/s11207-009-9375-0.
33. Delannée, C.; Aulanier, G. Cme Associated with Transequatorial Loops and a Bald Patch Flare. Sol. Phys. 1999, 190, 107–129. doi:10.1023/A:1005249416605.
34. Chen, P.F.; Wu, S.T.; Shibata, K.; Fang, C. Evidence of EIT and Moreton Waves in Numerical Simulations. Astrophys. J. Lett. 2002, 572, L99–L102. doi:10.1086/341486.
35. Chen, P.F.; Fang, C.; Shibata, K. A Full View of EIT Waves. Astrophys. J. 2005, 622, 1202–1210. doi:10.1086/428084.
36. Chandra, R.; Chen, P.F.; Fulara, A.; Srivastava, A.K.; Uddin, W. Peculiar Stationary EUV Wave Fronts in the Eruption on 2011 May 11. Astrophys. J. 2016, 822, 106. doi:10.3847/0004-637X/822/2/106.
37. Priest, E.R.; Démoulin, P. Three-dimensional magnetic reconnection without null points. 1. Basic theory of magnetic flipping. J. Geophys. Res. 1995, 100, 23443–23464. doi:10.1029/95JA02740.
38. Démoulin, P.; Priest, E.R.; Lonie, D.P. Three-dimensional magnetic reconnection without null points 2. Application to twisted flux tubes. J. Geophys. Res. 1996, 101, 7631–7646. doi:10.1029/95JA03558.
39. Zong, W.; Dai, Y. Mode Conversion of a Solar Extreme-ultraviolet Wave over a Coronal Cavity. Astrophys. J. Lett. 2017, 834, L15, doi:10.3847/2041-8213/834/2/L15.
40. Chen, P.F.; Fang, C.; Chandra, R.; Srivastava, A.K. Can a Fast-Mode EUV Wave Generate a Stationary Front? Sol. Phys. 2016, 291, 3195–3206. doi:10.1007/s11207-016-0920-3.
41. Lemen, J.R.; Title, A.M.; Akin, D.J.; Boerner, P.F.; Chou, C.; Drake, J.F.; Duncan, D.W.; Edwards, C.G.; Friedlaender, F.M.; Heyman, G.F.; et al. The Atmospheric Imaging Assembly (AIA) on the Solar Dynamics Observatory (SDO). Sol. Phys. 2012, 275, 17–40. doi:10.1007/s11207-011-9776-8.
42. Pesnell, W.D.; Thompson, B.J.; Chamberlin, P.C. The Solar Dynamics Observatory (SDO). Sol. Phys. 2012, 275, 3–15. doi:10.1007/s11207-011-9841-3.
43. Schou, J.; Scherrer, P.H.; Bush, R.I.; Wachter, R.; Couvidat, S.; Rabello-Soares, M.C.; Bogart, R.S.; Hoeksema, J.T.; Liu, Y.; Duvall, T.L.; et al. Design and Ground Calibration of the Helioseismic and Magnetic Imager (HMI) Instrument on the Solar Dynamics Observatory (SDO). Sol. Phys. 2012, 275, 229–259. doi:10.1007/s11207-011-9842-2.

44. Gopalswamy, N.; Yashiro, S.; Michalek, G.; Stenborg, G.; Vourlidas, A.; Freeland, S.; Howard, R. The SOHO/LASCO CME Catalog. Earth Moon Planets 2009, 104, 295–313. doi:10.1007/s11038-008-9282-7.
45. Moore, R.L.; Hagyard, M.J.; Davis, J.M.; Porter, J.G. The MSFC Vector Magnetograph, Eruptive Flares, and the SOLAR-A X-ray Images. In Flare Physics in Solar Activity Maximum 22; Uchida, Y., Canfield, R.C., Watanabe, T., Hiei, E., Eds.; 1991; Volume 387, p. 324. doi:10.1007/BFb0032658.
46. Romano, P.; Zuccarello, F.P.; Guglielmino, S.L.; Zuccarello, F. Evolution of the Magnetic Helicity Flux during the Formation and Eruption of Flux Ropes. Astrophys. J. 2014, 794, 118. doi:10.1088/0004-637X/794/2/118.
47. Zuccarello, F.P.; Chandra, R.; Schmieder, B.; Aulanier, G.; Joshi, R. Transition from eruptive to confined flares in the same active region. Astron. Astrophys. 2017, 601, A26, doi:10.1051/0004-6361/201629836.
48. Veronig, A.M.; Temmer, M.; Vršnak, B. High-Cadence Observations of a Global Coronal Wave by STEREO EUVI. Astrophys. J. Lett. 2008, 681, L113, doi:10.1086/590493.
49. Li, T.; Zhang, J.; Yang, S.; Liu, W. SDO/AIA Observations of Secondary Waves Generated by Interaction of the 2011 June 7 Global EUV Wave with Solar Coronal Structures. Astrophys. J. 2012, 746, 13, doi:10.1088/0004-637X/746/1/13.
50. Olmedo, O.; Vourlidas, A.; Zhang, J.; Cheng, X. Secondary Waves and/or the "Reflection" from and "Transmission" through a Coronal Hole of an Extreme Ultraviolet Wave Associated with the 2011 February 15 X2.2 Flare Observed with SDO/AIA and STEREO/EUVI. Astrophys. J. 2012, 756, 143, doi:10.1088/0004-637X/756/2/143.
51. Schmidt, J.M.; Ofman, L. Global Simulation of an Extreme Ultraviolet Imaging Telescope Wave. Astrophys. J. 2010, 713, 1008–1015. doi:10.1088/0004-637X/713/2/1008.
52. Piantschitsch, I.; Vršnak, B.; Hanslmeier, A.; Lemmerer, B.; Veronig, A.; Hernandez-Perez, A.; Calogovi´c, J.; Žic, T. A Numerical ˇ Simulation of Coronal Waves Interacting with Coronal Holes. I. Basic Features. Astrophys. J. 2017, 850, 88. doi:10.3847/1538-4357/aa8cc9.